\documentclass[prb,amsmath,amssymb,twocolumn]{revtex4-1}
\usepackage{graphicx}
\usepackage{epsfig}
\usepackage{dcolumn}
\usepackage{bm}
\usepackage{rotating}
\usepackage[table]{xcolor}
\usepackage[english]{babel}

\begin{document}
\widetext 

\title{The GW plus cumulant method and plasmonic polarons: application to the homogeneous electron gas}

\author{Fabio Caruso}
\author{Feliciano Giustino}
\affiliation{Department of Materials, University of Oxford, Parks Road, Oxford OX1 3PH, United Kingdom} 
\date{\today}

\date{\today}
\begin{abstract}
We study the spectral function of the homogeneous electron gas 
using many-body perturbation theory and the cumulant expansion.
We compute the angle-resolved spectral function 
based on the $GW$ approximation and the `$GW$ plus cumulant' approach. 
In agreement with previous studies, the $GW$ spectral function exhibits 
a spurious plasmaron peak at energies 1.5$\omega_{\rm pl}$ below the 
quasiparticle peak, $\omega_{\rm pl}$ being the plasma energy.
The $GW$ plus cumulant approach, on the other hand, reduces significantly the 
intensity of the plasmon-induced spectral features and renormalizes their energy 
relative to the quasiparticle energy to $\omega_{\rm pl}$.
Consistently with previous work on semiconductors, 
our results show that the HEG is characterized by the emergence of plasmonic 
polaron bands, that is, broadened replica of the quasiparticle bands, red-shifted 
by the plasmon energy. 
\end{abstract}
\maketitle

\section{Introduction}

The homogeneous electron gas (HEG) denotes a model system of electrons
interacting with a compensating homogeneous positively-charged background.\cite{nozieres1999theory}
The model, because of its simplicity, lends itself to analytical treatment
and has therefore provided the ideal test-case for the early development 
of many-body perturbation theory.\cite{gross1991many}
Despite the simplicity of the model, the HEG provides valuable insight into 
the physical properties of real systems, such as crystalline solids, as it 
exhibits prototypical features induced by electronic correlation. 
For example, the dielectric function of the HEG exhibits signatures of  
collective charge-density fluctuations, that is plasmons,\cite{pines1956} and the study of 
these features has led to the interpretation of the satellite structures 
in the early electron-energy loss spectroscopy (EELS) 
measurements of simple metals.\cite{Blackstock1955}
Overall, the study of plasmons has played an important role in the early development 
of many-body perturbation theory. 
The random-phase approximation (RPA), for example, was originally introduced 
by Pines and Bohm\cite{pines1952} as a simplification of the equation of motion 
for the density fluctuation in the HEG.

In the context of spectroscopy the study of plasmon-induced signatures 
in the spectral function of the HEG has contributed 
(i) to elucidate the fundamental processes that underpin the 
emergence of satellites in photoelectron spectra\cite{Lundqvist1967,Hedin19701}, and (ii) 
to derive new theoretical tools for their description.\cite{Langreth1970} 
Calculations based on the $GW$ approximation\cite{Hedin1965,hybertsenlouie1986} 
do not generally provide an 
accurate description of plasmon-induced spectral features. 
Both in the HEG\cite{Lundqvist1967/2,Lundqvist1968} and in
real solids,\cite{Kheifets2003,guzzo/2011} (for example silicon) 
the $GW$ approximation introduces a spurious 
`plasmaron' peak in the spectral function, that is a sharp quasiparticle-like
feature that arises from an additional solution of the Dyson equation. 
At first, the plasmaron was attributed to a novel type of quasiparticle excitation
resulting from the strong coupling between electrons and plasmons.\cite{Lundqvist1967/2,Lundqvist1968}
Later studies revealed that the plasmaron solution is an artifact of the
$GW$ approach, and disappears when a higher level of theory is employed, such as the 
cumulant expansion.\cite{Langreth1970}
The cumulant expansion approach 
is the state-of-the-art technique for the description of satellites in photoemission and it
accounts for the interaction between electrons and plasmons 
employing an independent boson model.\cite{Mahan2000} This model is exactly solvable 
for a single core electron interacting with a plasmon bath, and it provides
an explicit expression for the spectral function.\cite{Langreth1970,hedin1980}
Beside the first cumulant studies of the HEG, 
the cumulant expansion has been extended\cite{aryasetiawan/1996,hedin1999} and applied 
to describe the spectral signatures of 
plasmons in the valence photoelectron spectra of metals,\cite{aryasetiawan/1996,guzzo2014,sky2015} 
semiconductors,\cite{guzzo/2011,guzzo2012,lischner2013,caruso/2015/prl,caruso/2015/prb} 
and models systems.\cite{kas/2014,lischner/2deg}
The cumulant approach proved useful also in the computation 
of total energies\cite{galitskiimigdal} and ultrafast quasiparticle 
dynamics.\cite{Gumhalter2005,Gumhalter2012,silkin2015}
\begin{figure*}
\includegraphics[width=0.9\textwidth]{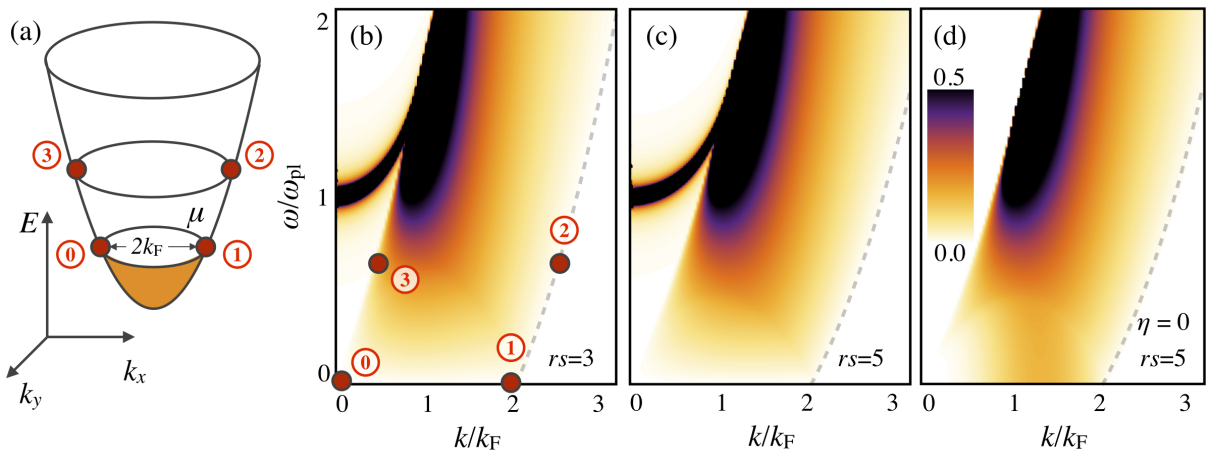}
\caption{
Schematic representation of the parabolic band dispersion of the HEG. The red circles (`1', `2', and `3') 
indicate some representative final states for electron-hole excitations starting from the initial 
state `0' at the Fermi energy $\mu$. 
(b)-(c) RPA loss function of the HEG evaluated in the $GW$ approximation at 
the densities $r_s=3$ (b) and $r_s=5$ (c). Here we considered 
frequencies with a small imaginary component ($\eta\sim0.004\times\mu$) to clearly visualize the 
plasmon peak. The case $\eta=0$ is shown in (d) for $r_s=5$.
Energy and momentum are expressed in units of the plasmon
energy $\omega_{\rm pl}$ and Fermi momentum $k_{\rm F}$, respectively.
The circles (0, 1, 2, and 3) in panel (b) are placed at points where the 
intensity of the loss function arises from the transitions
illustrated in panel (a).
}
\label{fig:EELS} 
\end{figure*}

In this work, we present a study of the spectral function 
and the signatures of electron-plasmon interaction in the HEG
based on the $GW$ approximation and the $GW$ plus cumulant ($GW+$C) approach. 
We first review the characterization of electronic excitations through the 
computation of the RPA dielectric function. 
We thus compute the angle-resolved spectral function of the HEG
in the $GW$ approximation to illustrate the emergence of the
spurious plasmaron peak.
Finally, we present calculations of HEG spectral function based on the
$GW$+C approach. Our calculations show that the $GW$+C approach 
renormalizes the energy of the plasmon-induced spectral features to 
$\sim\omega_{\rm pl}$ below the quasiparticle energy, consistently with previous work.  
Additionally, the analysis of the energy-momentum dispersion relations reveals the 
emergence of a plasmonic polaron band, which manifests itself as a broadened 
replica of the quasiparticle band, red-shifted by the plasmon energy.
This result further validates the concept of plasmonic polaron 
band, originally proposed for simple semiconductors\cite{caruso/2015/prl,caruso/2015/prb,lischner2015} 
and confirmed through angle-resolved photoemission measurements in silicon.\cite{lischner2015}

The manuscript is organized as follows. 
In Sec.~\ref{sec:rpa}, we present calculations of the loss function of
the HEG in the RPA. The $GW$ approximation and its application to the 
spectral properties of the HEG are discussed in Sec.~\ref{sec:gw}, whereas 
in Sec.~\ref{sec:gwpc} we present calculations of the HEG spectral function 
based on the $GW+$C approach. 
Finally, our conclusions are presented in Sec.~\ref{sec:conc}.

\section{Signatures of plasmons in the dielectric function}\label{sec:rpa}

Electronic excitations of the HEG can be characterized 
through the computation of the loss function:\cite{Nozieres1959}
\begin{align}\label{eq:loss}
\mathcal{L}({\bf q},\omega)= {\rm Im}\, \epsilon^{-1}({\bf q},\omega), 
\end{align}
where $\epsilon$ is the dielectric function.
Since the dielectric function vanishes at the frequencies 
resonant with the excitations of plasmons,\cite{fetter} 
the loss function exhibits pronounced 
singularities at the plasmon energies $\omega_{\rm pl}({\bf q})$.
The condition $\epsilon[{\bf q},\omega_{\rm pl}({\bf q})]=0$,
\footnote{for ${\bf q}=0$ this condition yields the plasma frequency 
$\omega_{\rm pl} = \omega_{\rm pl}({\bf q}=0)=\sqrt{4\pi n e^2/m}$,  $n$ being the 
HEG density} 
which defines the plasmon energy, provides a rational to distinguish between 
spectral signatures of plasmon and electron-hole pairs in the loss function. 
In particular, plasmons are expected to induce Dirac-delta-like 
features in Eq.~(\ref{eq:loss}), well separated from the continuum of
electron-hole pair excitations. 

\begin{figure*}
\includegraphics[width=0.75\textwidth]{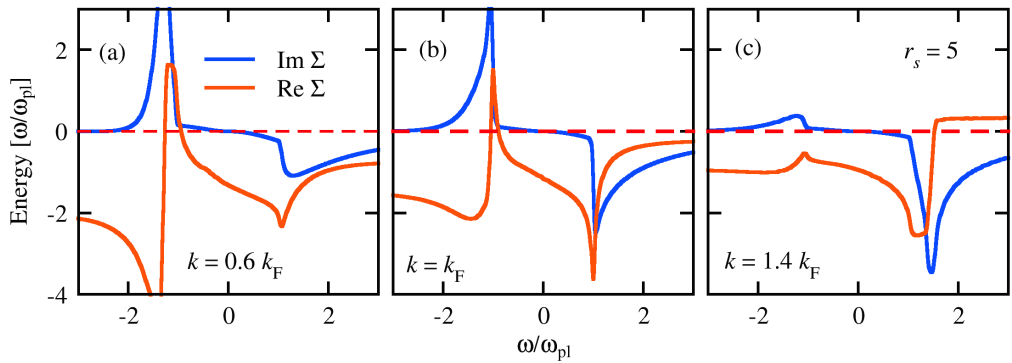}
\caption{$GW$ self-energy of the HEG for $r_s = 5$  at $k=0.6 k_{\rm F}$ (a),
$k=k_{\rm F}$ (b), and $k=1.4 k_{\rm F}$ (c). 
The energy is in units of the plasmon energy, 
$\omega_{\rm pl}=4.2$~eV.} 
\label{fig:Sgw} 
\end{figure*}

In a Green's function formalism, the dielectric function may be 
expressed as: 
\begin{align}\label{eq:eps0}
\epsilon({\bf q},\omega) = 1-v({\bf q})\chi_0({\bf q},\omega), 
\end{align}
where we introduced the irreducible polarizability $\chi_0$ and the 
bare Coulomb interaction $v({\bf q})=4\pi/| {\bf q}| ^2$.
Here and in the following we adopted Hartree atomic units, unless otherwise stated.
In the RPA, whereby electron-hole interactions are neglected, 
$\chi_0$ may be expressed explicitly as:\cite{Hedin19701}
\begin{align}\label{eq:chi0}
\chi_0({\bf q},\omega) = -i \int \frac{d{\bf k}\,d\omega'}{(2\pi)^4} G_0({\bf q}+{\bf k},\omega+\omega') G_0({\bf k},\omega').
\end{align}
We introduced here the non-interacting 
Green's function, defined by:
\begin{align}
G_0({\bf k},\omega) = \frac{1}{ \omega +\mu - \epsilon_{\bf k} + 
i {\rm sign (\mu - \epsilon_{\bf k})}},
\end{align}
where $\mu$ is the Fermi energy, $\epsilon_{\bf k}=k^2/2$, and $\eta$ a positive infinitesimal.
The convolution in Eq.~(\ref{eq:chi0}) may be carried out analytically,\cite{fetter}
yielding an explicit expression for the dielectric function of the HEG:\cite{Hedin1965,Mahan2000}
\begin{align}\label{eq:eps}
\epsilon(q,\omega)=1+\frac{\alpha r_s}{8\pi q^3}[H(q+u/q)-H(q-u/q)],
\end{align}
where $H(q) = 2q+(1-q^2){\rm ln}\left[ (q+1)/(q-1)\right]$, $\alpha=(4/9\pi)^{1/3}$. 
Here we followed the notation of Ref.~\onlinecite{Hedin1965}  where 
$q$ denotes momenta in units of $2k_{\rm F}$ and $u$ are energies in units of 
$4\mu$. $r_s$ denotes the Wigner-Seitz radius.
The calculation of the loss function is reduced to the evaluation 
of the Eq.~(\ref{eq:eps}) for several frequencies and momenta. 

In Fig.~\ref{fig:EELS} we report the loss function of the HEG for 
$r_s=3$ (b) and $r_s=5$ (c). 
A detailed discussion of the loss function may be found 
in many textbooks.\cite{Mahan2000,pines1999elementary} 
Briefly, the broad band of width $~2k_{\rm F}$ is the 
continuum of electron-hole excitation. 
To exemplify the origin of these features we report in Fig.~\ref{fig:EELS}~(a)
a schematic representation of the free electron energy band of the HEG.
The labels `1', `2', and `3' denote possible final states for 
the excitation of an electron at the Fermi energy (red dot labelled~`0'). 
The transition to 1 involves a very small change of energy and a momentum transfer of 
$\sim 2 k_F$. 
The contribution of these transition to the loss function is infinitesimal due to the 
small phase space available for the transitions. 
By considering finite energy changes, electron-hole excitations must necessarily 
involve a change of momentum. The maximum  (minimum) momentum transfer would 
correspond to transition of the type 2 (3). The corresponding signatures of these transitions 
in the loss function are indicated by 1, 2, and 3 in Fig.~\ref{fig:EELS}~(b).
According to the previous discussion, the high-intensity feature at small momentum transfer 
may not be attributed to electron-hole excitations. 
This feature is the plasmon peak and stems from the zeros of dielectric function. 
In particular, for $k=0$, the plasmon peak occurs exactly at the plasma energy~$\omega_{\rm pl}$. 
At energies and momenta at which the plasmons and the 
electron-hole excitations coexist, the plasmon peak is broadened out and its spectral features
are not distinguishable from the electron-hole continuum.

To visualize the plasmon peak of the loss function we considered frequencies with a 
small imaginary part ($\eta=0.004\times\mu$). 
If purely real frequencies were considered,  
the plasmon peak would not be visible
in the loss function owing to the finite momentum resolution in the figure 
[Fig.~\ref{fig:EELS}~(d), for  $r_s=5$]. 
We now move on to discuss the spectral function of the HEG in the $GW$ approximation and 
the spectral signatures of plasmons.

\section{$GW$ self-energy and spectral function of the HEG}\label{sec:gw}

In the $GW$ approximation, the electron self-energy for the HEG takes the form:
\begin{align}\label{eq-GW}
\Sigma({\bf k},\omega) = \frac{i}{2\pi} 
\int d\omega d{\bf q} G({\bf k+q},\omega+\omega') W({\bf q},\omega) .
\end{align}
The screened Coulomb interaction $W$ can be expressed as:
\begin{align}\label{eq-W}
W({\bf q},\omega)=\frac{v({\bf q})}{\epsilon({\bf q},\omega )} \quad. 
\end{align}
In principle, the evaluation of the self-energy in Eq.~(\ref{eq-GW}) should employ
a Green's function obtained self-consistently from the solution of the 
Dyson's equation:
\begin{align}\label{eq:dyson}
[G({\bf k},\omega)]^{-1}= [G_0({\bf k},\omega)]^{-1} - \Sigma({\bf k},\omega).
\end{align}
Self-consistent $GW$ denotes the procedure in which Eqs.~(\ref{eq:eps0}), 
(\ref{eq:chi0}), (\ref{eq-GW}), (\ref{eq-W}), and (\ref{eq:dyson}) are 
iterated until convergence is reached.
For atoms and  molecules it is well established that 
self-consistent $GW$ improves the description of quasiparticle 
energies\cite{stan/2006,stan,thygesen,marom/2012/prb,caruso/2013/prb,koval/2014} 
as compared to non-self-consistent calculations. 
Similar conclusions have been obtained for the total energies of atoms,\cite{dahlenleeuwen2005,dahlenleeuwen2006,stan} 
molecules,\cite{caruso/2012/prb,caruso/2013/prl,hellgren/2015/prb} and the 
homogeneous electron gas.\cite{holmvonbarth1998,holm1999,galitskiimigdal} 
For what concerns plasmon satellites in the spectral function, however, 
Holm and Von Barth have shown that 
self-consistent $GW$ deteriorates the spectral function 
due to a spurious renormalization of the satellite intensity.\cite{holmvonbarth1998}
In the following we will limit the discussion to 
`one-shot' $GW$ (or $G_0W_0$), in which Eq.~(\ref{eq-GW}) is evaluated at the 
first-iteration of the self-consistent procedure.
The $GW$ self-energy [Eq.~(\ref{eq-GW})] 
has been obtained from the numerical integration Eq.~(89)-(91) of Ref.~\onlinecite{Hedin1965}. 
In Fig.~\ref{fig:Sgw} we illustrate the real and imaginary part of the self-energy (in units of the plasmon energy 
$\omega_{\rm pl}$)  for $r_s = 5$. 
These results, based on the calculation of the RPA dielectric function, 
are in excellent agreement with the results reported by Lundqvist based on the 
plasmon-pole approximation.\cite{Lundqvist1967} 
The $GW$ self-energy exhibits a sharp pole at the energy 
$\epsilon_{\bf k}\pm\omega_{\rm pl}$, where the $+/-$ signs hold for 
empty/occupied states. It is evident from Eqs.~(\ref{eq-GW})  and (\ref{eq-W}) that this 
feature stems primarily from the plasmon peak in the dielectric function, which introduces a singularity in 
the screened Coulomb interaction $W$ owing to the vanishing $\epsilon$.

Having reviewed the self-energy 
in the $GW$ approximation, we move now to discuss the signatures of 
plasmon excitations in the spectral function of the HEG. 
\begin{figure*}
\includegraphics[width=0.98\textwidth]{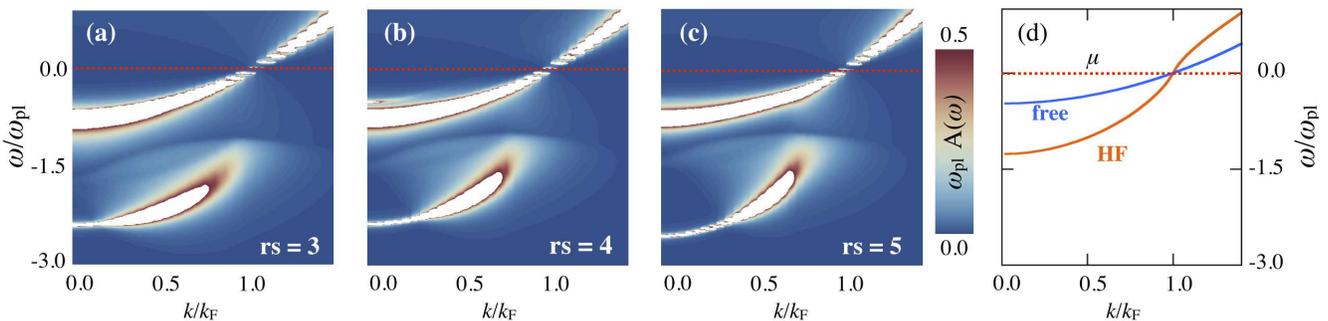}
\caption{Spectral function of the HEG evaluated in the $GW$ 
approximation for (a) $r_s = 3$, (b) $r_s=4$, and (c) $r_s=5$. 
Energy and momentum are expressed in units of the plasmon 
energy $\omega_{\rm pl}$ and Fermi momentum $k_{\rm F}$, respectively. 
To facilitate a comparison on the same scale, the intensity of the spectral function 
has been multiplied by $\omega_{\rm pl}$. 
The red dotted lines indicate the Fermi energy. 
(d) Quasiparticle energy versus momentum dispersion relations within the Hartree-Fock (HF) 
approximation, and for non-interacting electrons (free).}
\label{fig:Agw} 
\end{figure*}
The spectral function is given by:
  \begin{align}\label{eq:specfun}
  A({\bf k},\omega)
  = \frac{1}{\pi} \frac{|\Sigma^{\prime\prime}_{{\bf k}}(\omega)|}
  {[\omega - \epsilon_{{\bf k}} - \Sigma^\prime_{{\bf k}}(\omega)]^2
  + [\Sigma^{\prime\prime}_{{\bf k}}(\omega)]^2} ,
  \end{align}
where $\Sigma^\prime$ and $\Sigma^{\prime\prime}$ denote the real and imaginary part of the 
$GW$ self-energy, respectively. 
In independent-particle approximations, 
such as the Hartree-Fock approximation or Kohn-Sham density functional theory, 
the self-energy is static (independent of frequency) and real. 
The spectral function reduces to a Dirac delta function:
\begin{equation}\label{eq:specfun-hf}
  A({\bf k},\omega) = \delta(\omega - \epsilon_{{\bf k}} - \Sigma_{\bf k}).
  \end{equation}
In the case of the Hartree-Fock approximation, the self-energy may be obtained analytically,\cite{martin2004electronic} and the evaluation of the spectral function
is straightforward.
The Hartree-Fock spectral function is reported in Fig.~\ref{fig:Agw}(d), 
 alongside with the free-electron spectral function.
At each ${\bf k}$ point, the spectral function of the HEG 
exhibits Dirac-delta-like structures at the energy of quasiparticle excitations. 
However, there are no structures that may be attributed to collective excitations induced 
by electronic correlation. 
In the $GW$ approximation, on the other hand, the self-energy is characterized by a complex 
frequency dependence which introduces several additional signatures of electron correlation 
in the spectral function of the HEG.

The $GW$ spectral function is illustrated in Fig.~\ref{fig:Agw} for the HEG at three different 
electron densities ($r_s = 3 $, $4$, and $5$). 
The quasiparticle band is the bright band that appears at $\omega\simeq-\omega_{\rm pl}$ for ${\bf k} = 0$ 
and increases quadratically with momentum. 
At variance with the independent particle approximation, the quasiparticle peaks 
acquire a broadening (vanishing at the Fermi energy) which stems from electronic 
correlation and is related to the finite lifetime of electronic excitations. 

Beside the quadratic quasiparticle band, the spectral function presents pronounced 
spectral features at energies $1.5\, \omega_{\rm pl}$ below the quasiparticle energy. 
These features are additional solutions of the quasiparticle equation, that is, 
they arise from the zeros of $\omega - \epsilon_{{\bf k}} - \Sigma^\prime_{{\bf k}}(\omega)$ 
in Eq.~(\ref{eq:specfun}).
These spectral features, first reported by Lundqvist, 
have been originally attributed to plasmarons, 
a new type of quasiparticle stemming from the strong 
coupling between holes and plasmons.\cite{Lundqvist1967/2,Lundqvist1968}
However, subsequent work have shown that plasmarons are an
artifact of the $GW$ approximation.\cite{Langreth1970,guzzo/2011}
%
%As alluded to above, self-consistency deteriorates  
%the description of satellites even further as compared to one-shot $G_0W_0$
%calculations.\cite{holmvonbarth1998} 
%The inclusion of beyond-$GW$ Feynman diagrams through vertex corrections 
%in the self-energy would provide a rigorous and systematic way to improve 
%the spectral properties of the HEG.\cite{ferdi1998} 
%However, %to avoid the additional computational cost required by vertex correction,  
%vertex corrections increase significantly the numerical 
%complexity of $GW$ calculations and simple static vertex 
%corrections  do not yield significant improvement of 
%the spectral properties.
As shown in the following, the cumulant expansion approach provides an ideal 
way to address this problem, as it improves the plasmon-induced spectral features 
of the HEG at the same computational cost of a $GW$ calculation. 

\begin{figure}
\includegraphics[width=0.38\textwidth]{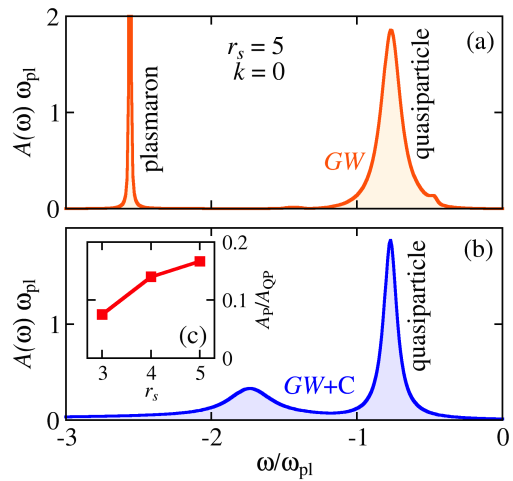}
\caption{Spectral function of the HEG evaluated from $GW$ (a) and $GW$+C (b)
for ${\bf k}=0$ at $r_s=5$. (c) Ratio between the intensity of the plasmon satellite and the quasiparticle 
peak (as obtained form $GW$+C) as a function of $r_s$.}
\label{fig:gamma}
\end{figure}
\begin{figure*}
\includegraphics[width=0.8\textwidth]{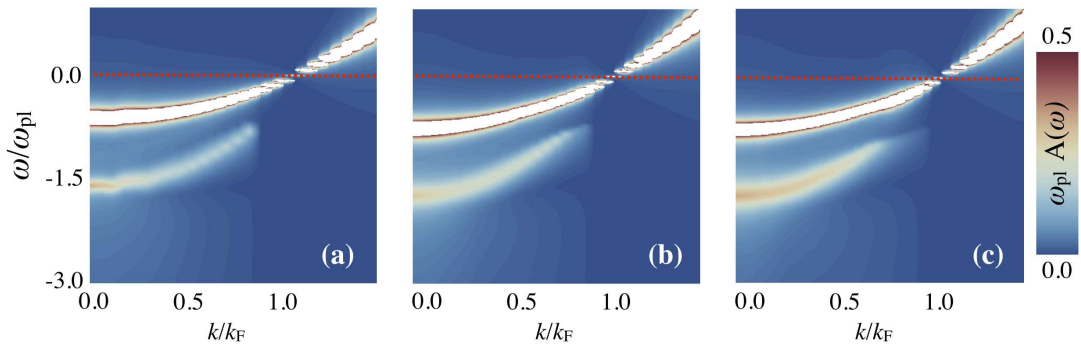}
\caption{Spectral function of the HEG evaluated from the $GW$+C
approach for (a) $r_s = 3$, (b) $r_s=4$, and (c) $r_s=5$. 
Energy and momentum are expressed in units of the plasmon 
energy $\omega_{\rm pl}$ and Fermi momentum $k_{\rm F}$, respectively. 
To facilitate a comparison on the same scale, the spectral function 
intensity has been multiplied by $\omega_{\rm pl}$. 
The dashed lines indicate the Fermi energy.}
\label{fig:Agwc} 
\end{figure*}

\section{Spectral function from the $GW$ plus cumulant approach}\label{sec:gwpc}

For the computation of spectral properties, it is common practice to 
combine the cumulant expansion with the $GW$ approximation. 
In the resulting $GW+$C approach, the spectral function can
be expressed as:\cite{aryasetiawan/1996}
  \begin{equation}\label{eq-spectrum1}
  A({\bf k},\omega) =  [ A^{\rm QP}({\bf k},\omega) + 
  A^{\rm QP}({\bf k},\omega)\ast A^{\rm C}({\bf k},\omega)]. 
  \end{equation}
This expression corresponds to the first-order cumulant expansion, 
and it ignores processes in which multiple plasmons are excited. Multi-plasmon processes
may be accounted for by including higher-order 
cumulant terms.\cite{guzzo/2011,kas/2014,sky2015} 
The first term in Eq.~(\ref{eq-spectrum1}) is the ordinary quasiparticle spectral function, 
defined as:
\begin{align}\label{eq.specfun}
A^{\rm QP}({\bf k},\omega)  
= \frac{1}{\pi} \frac{|\Sigma''_{{\bf k}} (\epsilon^{\rm qp}_{{\bf k}})|}
{[\omega - \epsilon_{{\bf k}} - \Sigma'_{{\bf k}}(\epsilon^{\rm qp}_{{\bf k}})]^2 
+ [\Sigma''_{{\bf k}}(\epsilon^{\rm qp}_{{\bf k}})]^2},
\end{align}
%
%where $\Sigma_{\bf k}$ denotes the $GW$  self-energy and  $\epsilon_{{\bf k}}={\bf k}^2/2$. 
%
%As discussed in Ref.~\onlinecite{Caruso2015},
%the convolution product in the second term of Eq.~(\ref{eq-spectrum1}) introduces novel dispersive features
%in the spectral function, which
%account for events in which a photo-hole and a plasmon are excited simultaneously.
$\epsilon^{\rm qp}_{{\bf k}} = \epsilon_{{\bf k}} + \Sigma'_{{\bf k}}(\epsilon^{\rm qp}_{{\bf k}}) $ 
is the quasiparticle energy. 
The term $A^{\rm C}$ is defined as:
  \begin{equation}\label{eq-spectrum2}
  A^{\rm C}({\bf k},\omega)  
  =  \frac{\beta_{{\bf k}}(\omega) - \!\beta_{{\bf k}}(\epsilon_{{\bf k}}) - 
  \!(\omega-\epsilon_{{\bf k}})\!\left. \displaystyle\frac{\partial \beta_{{\bf k}}}{\partial \omega}
    \right|_{\epsilon_{{\bf k}}}}
  {(\omega-\epsilon_{{\bf k}})^2},
  \end{equation}
where $\beta_{{\bf k}}(\omega) = \pi^{-1}{\rm Im}\Sigma_{{\bf k}}(\epsilon_{{\bf k}}-\omega)
\theta(\mu-\omega)$.
Equation~(\ref{eq.specfun}) accounts for the contribution of quasiparticle 
excitations to the spectral functions in absence of plasmons.
The second term in Eq.~(\ref{eq-spectrum1}) accounts for processes in which 
an electron is emitted and a plasmon is excited. 

We evaluated the $GW$+C angle-resolved 
spectral function of the HEG by combining the $GW$ self-energy defined in Eq.~(\ref{eq-GW}) 
with the cumulant expansion defined by Eqs.~(\ref{eq-spectrum1})-(\ref{eq-spectrum2}).
In Fig.~\ref{fig:gamma} we compare the spectral function at $r_s=5$ and ${\bf k} =0$ obtained 
from $GW$ (a) and $GW$+C (b). 
The $GW$ and the $GW$+C approaches provide a similar description of the quasiparticle peak. 
In the $GW$+C approach, these features stems from 
Eq.~(\ref{eq.specfun}) which coincides with the $GW$ 
spectral function [Eq.~(\ref{eq:specfun})] at 
the quasiparticle energy $\epsilon^{\rm qp}_{{\bf k}}$. 
The changes introduced by the cumulant approach affect 
primarily the low-energy part of the spectral function.
At variance with the $GW$ spectral function, whereby the plasmon peak is red-shifted by 
approximately $1.5\,\omega_{\rm pl}$ with respect to the quasiparticle band, 
the $GW$+C yields a satellite structure separated by $\sim\omega_{\rm pl}$ 
from the quasiparticle energy.
As compared to the $GW$ spectral function, these 
spectral features are more broad and less intense.

Inspecting the angle-resolved spectral function, shown in 
Fig.~~\ref{fig:Agwc} for (a) $r_s=3$, (b) $r_s=4$, and (c) $r_s=5$,
we note that the dispersion of $GW$+C satellite 
follows closely the momentum dependence of the quasiparticle bands. 
This indicates that also the HEG is characterized by
the formation of a well-defined plasmonic polaron band.
Plasmonic polaron bands are a manifestation of
the simultaneous excitation of a hole (for instance via
the absorption of a photon) and the excitation of a plasmon, 
and they manifest themselves as broadened band-structure replica, 
shifted by the plasmon energy with respect to the ordinary quasiparticle bands. 
These spectral features have recently been predicted in the 
context of $sp$-bonded semiconductors\cite{caruso/2015/prl,caruso/2015/prb} 
and confirmed by angle-resolved photoemission spectroscopy measurements of silicon.\cite{lischner2015}
In the case of silicon, plasmonic polaron bands replicate the entire set of valence bands.
The HEG, on the other hand, is characterized by a single band. Correspondingly, a single 
plasmonic polaron band can be observed in Fig.~\ref{fig:Agwc}.

Our calculations show that the intensity of the satellite
features in the  $GW$+C spectral function
decreases with increasing density (that is with decreasing $r_s$), as shown in Fig.~\ref{fig:gamma}(c).
This behaviour may be attributed to the different scaling of the 
Coulomb interaction and the kinetic energy with the changes of the electron density:\cite{nozieres1999theory}
at large densities, the kinetic energy increases more rapidly than the Coulomb interaction and, 
correspondingly, the effect of electron correlation becomes less important as compared to the kinetic term. 
In the limit of infinite electron density, the HEG can 
be approximately described by a non-interacting HEG, 
as the Coulomb interaction becomes negligible, and the satellite is expected to disappear completely.
%At high densities, the Coulomb interaction $U$ between charge carriers becomes weak 
%as compared to the kinetic energy $K$. In particular, 
%one may show that $U/K\sim r_s$. 
%The electron behaviour is thus dominated by the kinetic energy 
%and the system becomes only weakly correlated. 
%In particular, in the high-density limit ($r_s\ll 1$) Coulomb interactions can be ignored,  
%the HEG may approximately be described as a non-interacting particle systems and, correspondingly, 
%the satellite is expected to disappear. 
Conversely, the Coulomb interaction dominates at low densities (large $r_s$) 
and one may expect a more pronounced effect of electron correlation on the spectral properties.

\section{Summary and conclusions}\label{sec:conc}

In summary, we have presented a study of 
spectral function of the homogeneous electron gas, with an emphasis on the
signatures of electron-plasmon interactions. 
In particular, we reviewed the analysis of  the loss 
function of the HEG in the random phase approximation and computed 
the spectral function of the HEG from the $GW$ approximation and the $GW$+C approach.

At variance with calculations in the independent-particle approximation, 
the explicit treatment of electron-electron interaction
within the $GW$ approximation introduces a non-trivial 
frequency dependence which, in turn, leads to the emergence of 
additional low-energy features in the spectral function. 
%These spectral features may be attributed to the 
At the $GW$ level, for $k<k_{\rm F}$ the spectral function exhibits
the spurious plasmaron peak 
at an energy of approximately $1.5~\omega_{\rm pl}$ below the quasiparticle energy. 
A more advanced description of electron-plasmon coupling within the $GW$+C approach, however,  
reduces significantly the intensity of the plasmon-induced spectral features and
renormalizes their energy difference to the quasiparticle band to the 
plasma energy $\omega_{\rm pl}$.
Consistently with previous work on semiconductors, 
the present study reveals that also the HEG is characterized by the emergence 
of plasmonic polaron bands, that is, plasmon-induced band structure 
replica red-shifted by the plasmon energy. 

\acknowledgements
This work was supported by the Leverhulme Trust (Grant No. RL-2012-001) 
and the European Research Council (EU FP7/ERC Grant No. 239578 and EU FP7/Grant 
No. 604391 Graphene Flagship). Calculations were performed at the Oxford 
Supercomputing Centre\cite{ARC} and at the Oxford Materials Modelling Laboratory.

\bibliography{references}

\end{document}